\documentclass[a4paper,twoside]{article}

\usepackage{epsfig}
\usepackage{subcaption}
\usepackage{calc}
\usepackage{amssymb}
\usepackage{amstext}
\usepackage{amsmath}
\usepackage{amsthm}
\usepackage{multicol}
\usepackage{multirow}
\usepackage{pslatex}
\usepackage{url}
\usepackage{apalike}
\usepackage{bm}
\usepackage{xcolor}
\usepackage{subcaption}
\usepackage{SCITEPRESS}     

\begin{document}

\title{Longitudinal Evaluation of Open-Source Software Maintainability}
\author{
\authorname{Arthur-Jozsef Molnar and Simona Motogna}
\affiliation{Faculty of Mathematics and Computer Science, Babe\c{s}-Bolyai University, Cluj-Napoca, Romania}
\email{\{arthur, motogna\}@cs.ubbcluj.ro}
}


\abstract{We present a longitudinal study on the long-term evolution of maintainability in open-source software. Quality assessment remains at the forefront of both software research and practice, with many models and assessment methodologies proposed and used over time. Some of them helped create and shape standards such as ISO 9126 and 25010, which are well established today. Both describe software quality in terms of characteristics such as reliability, security or maintainability. An important body of research exists linking these characteristics with software metrics, and proposing ways to automate quality assessment by aggregating software metric values into higher-level quality models. We employ the Maintainability Index, technical debt ratio and a maintainability model based on the ARiSA Compendium. Our study covers the entire 18 year development history and all released versions for three complex, open-source applications. We determine the maintainability for each version using the proposed models, we compare obtained results and use manual source code examination to put them into context. We examine the common development patterns of the target applications and study the relation between refactoring and maintainability. Finally, we study the strengths and weaknesses of each maintainability model using manual source code examination as the baseline.}

\onecolumn \maketitle \normalsize \setcounter{footnote}{0} \vfill

\section{\uppercase{Introduction}}
\label{sec:introduction}
Maintainability represents the \textit{"degree of effectiveness and efficiency with which a product or system can be modified to improve it, correct it or adapt it to changes in environment, and in requirements"} \cite{23}.

We can identify many cases of large, complex software systems developed over a long period of time, during which they were subjected to a large number of modifications or extensions, many of which were not planned from project onset. These cases easily lead to increases in maintenance costs related with unchecked software complexity, leading to difficulties in localizing errors or the requirements for carrying out source code modifications. In many cases this is compounded by having different teams for development and maintenance, increasing program comprehension difficulty. This leads to maintenance being one of the key factors for continued increase of costs during the software lifecycle. 

One important cause is that maintenance is usually considered during the late stages of software development, which is usually focused on delivery. Providing the development team with adequate tool-backed methodologies facilitates foresight of difficult situations and can help avoid some of the technical problems that otherwise build up in what is now known as technical debt. Additional useful information is represented by discovering maintenance \textit{hotspots}, namely those parts of the source code that are most difficult to maintain. Even more, agile practices such as continuous delivery have reinforced the importance of integrating maintenance activities tightly within the development process. 

Our goal is to study the long term evolution of software maintainability in large open-source applications. We assess target application maintainability using relevant software quality models. Our focus on open-source software allows us to validate target application source code, check that all functionalities are present and apply proposed quality models. Several studies have investigated the relation between maintainability and software metrics \cite{7,8,21,22}, while others have investigated maintainability as a software quality factor \cite{12,23,24}. Previous work proves that a relation exists between software maintainability and metrics. However, further investigation is required to provide practitioners with actionable results that can be integrated into software development workflows.

The novelty of our approach consists in (i) studying the evolution of software maintainability over the long-term, by evaluating all released versions of the target applications using several quality models; (ii) identifying long-term trends in application maintainability that are common across the studied applications and (iii) studying the quality models themselves in order to analyze their strengths and weaknesses through the lens of complex software systems.

The following section is dedicated to detailing the maintainability models used in our study, while Section \ref{sec:CaseStudy} details our case study, carried out according to best practices \cite{17}. Relevant related work is presented in Section \ref{sec:RelatedWork}, while Section \ref{sec:Conclusion} is reserved for conclusions.

\section{\uppercase{Preliminaries}}
\label{sec:preliminaries}
As DeMarco underlines: \textit{"you cannot control what you cannot measure"} \cite{6}, software metrics have become essential indicators for controlling software. Starting with the earliest employed metrics, that measured the number of lines of code, functions, modules, or classes in a program, and continuing with metrics associated with complexity, they have proven their usefulness in measuring and controlling software development. This is especially true in the case of large-scale software. Object oriented metrics \cite{7} provided an important boost to the domain, by using metrics to detect code smells, improve maintainability and other software quality characteristics.

The importance of maintainability is illustrated by its inclusion as a quality factor in all proposed software quality models. The most recent standardized model is ISO 25010 \cite{23}, in which maintainability is comprised of sub-characteristics Modularity, Reusability, Analyzability, Modifiability and Testability. This provides an early indication regarding the aspects that are important to easily maintain code. Unfortunately it does not offer a methodology to improve this factor, nor a formalized way of assigning it a numerical value. To fulfill this need, several ways to model and measure maintainability have been proposed. We selected three such models for inclusion in our case study, and we describe them in the following sections.

\subsection{Maintainability Index}
\label{sec:MI}
The Maintainability Index (MI) was introduced in 1992 \cite{13} and is perhaps the most well established metric for software maintainability. The index can be applied at function, module or program level. While several variants of it exist, we employ a version that is detailed within several modern-day software tools \cite{14,32}:

$MI = 171 - 5.2 * ln(aveV) - 0.23 * aveG - 16.2 * ln(aveSTAT)$, where:
\begin{itemize}
\item \textit{aveV} - average Halstead volume. It reflects the computational load of the code in terms of operators and operands used.
\item \textit{aveG} - average cyclomatic complexity. It reflects the number of possible execution paths.
\item \textit{aveSTAT} - average number of source code statements. Variants exist using the number of lines of code, however the consideration is \cite{14} that the number of statements provides a better reflection of source code size.
\end{itemize}

In this form, MI values range from 171 down to negative numbers. However, negative MI values reflect very poor maintainability. As such, the following normalized value was introduced \cite{32}: $MI_{normalized} = max(0, \frac{100}{171}*MI)$. Normalized values range between 0 and 100, with values below 20 illustrating poor maintainability \cite{32}. Several code inspection and development tools can compute the MI, including JHawk \cite{14}, SonarQube \cite{15}, the Metrics .NET library \cite{16} and Microsoft Visual Studio \cite{32}. 

However, studies exist criticizing this metric \cite{8,9,19}. They illustrate different inconsistencies that occur when used in the context of current programming languages. In addition to various tools implementing different versions of the MI, such as adding an additional parameter for comments, we find disagreement regarding threshold values for poor maintainability \cite{14,20,32}. As the MI was proposed in the '90s, it does not consider object orientation, which emphasized new relations, such as inheritance, coupling and cohesion to the forefront. These were proven to have significant impact on maintainability \cite{7,22,21,5}.

\subsection{Technical Debt Ratio}
Technical debt provides a numerical representation of the required maintenance effort, tailored towards large applications. Technical debt is used increasingly often, especially in the context of agile development methodologies. Originally introduced as a metaphor borrowed from the financial sector \cite{29}, technical debt suggests that the debt accumulated during development is transformed in a financial debt to be paid in the context of release versions. Technical debt is considered a fair measure of the deficiencies in internal quality of software \cite{30}. The concept gained popularity after it was detailed in the Software Quality Assessment Based on Lifecycle Expectations (SQALE) methodology introduced by J.L. Letouzey \cite{31}. 

SonarQube is perhaps the most well known tool that calculates technical debt. It employs a configurable model where each issue discovered using static analysis is associated with an estimation of the time required to fix it. The technical debt represents the total time required to fix all discovered issues. The \textit{technical debt ratio} puts the debt into context, by dividing the amount of time required to fix all issues by the estimated time to create the software system\footnote{By default, SonarQube estimates 0.06 days to develop one line of code}. As such, technical debt ratio values range between 0 and 1. By default, the SonarQube model \cite{15} assigns the best maintainability rating of \textit{A} to code with technical debt ratio smaller than 0.05, where fixing all issues requires less than 5\% of the total development time. The worst rating of \textit{E} is assigned to code having debt ratio over 0.5.

\subsection{{ARiSA} Model for Maintainability}
\label{fig:ARISAModel}
Another approach for metric-driven measurement of software maintainability is based on the \textit{Compendium of Software Quality Standards and Metrics} \cite{12}, currently maintained by ARiSA\footnote{http://www.arisa.se/index.php?lang=en}, together with researchers at Linnaeus University\footnote{We refer to this work as the ARiSA Compendium}. The ARiSA Compendium explores the relation between software quality, as expressed by the ISO 9126 standard, and software metrics. The ISO 9126 quality model is comprised of six characteristics and 27 sub-characteristics. One of the six characteristics is \textit{Maintainability}, with sub-characteristics Analyzability, Changeability, Stability, Testability and Compliance. For each sub-characteristic, with the exception of Compliance, the Compendium provides a set of related metrics, as well as the strength and direction of each metric's influence. This information is synthesized in Table \ref{tab:ARISAMetrics}. As shown, the influence can be direct (plus sign, showing a direct relation between the metric value and corresponding sub-characteristic) or inverse (minus sign), as well as strong (doubled sign). For example, increased LOC has a strong and negative influence on Analyzability.

\begin{table}[t]
\caption{Metric influences on maintainability according to the {ARiSA} model}\label{tab:ARISAMetrics} \centering
\begin{tabular}{c c|c|c|c|c|}
  \cline{3-6} & & \multicolumn{4}{|c|}{Maintainability}\\ \cline{3-6}
  \raisebox{-1.75ex}{\rotatebox[origin=c]{90}{Category}} & \raisebox{-3ex}{\rotatebox[origin=c]{90}{Metric}} & \rotatebox[origin=c]{90}{ Analyzability} & \rotatebox[origin=c]{90}{ Changeability} & \raisebox{-2ex}{\rotatebox[origin=c]{90}{Stability}} & \raisebox{-1.25ex}{\rotatebox[origin=c]{90}{Testability}} \\ \hline
  \multirow{5}{*}{\rotatebox[origin=c]{90}{Complexity}} & \multicolumn{1}{|c|}{LOC} & $--$ & $--$ & - & $--$ \\ \cline{2-6}
  & \multicolumn{1}{|c|}{NAM} & $--$ & $--$ & - & $--$ \\ \cline{2-6}
  & \multicolumn{1}{|c|}{NOM} & $--$ & $--$ & - & $--$ \\ \cline{2-6}
  & \multicolumn{1}{|c|}{WMC} & $--$ & $--$ & - & $--$ \\ \cline{2-6}
  & \multicolumn{1}{|c|}{RFC} & $--$ & $--$ & - & $--$ \\ \hline
  \multirow{9}{*}{\rotatebox[origin=c]{90}{Structure}} & \multicolumn{1}{|c|}{DIT} & $--$ & $--$ & - & $--$ \\ \cline{2-6}
  & \multicolumn{1}{|c|}{NOC} & - & $--$ & - & - \\ \cline{2-6}
  & \multicolumn{1}{|c|}{CBO} & $--$ & $--$ & $--$ & $--$ \\ \cline{2-6}
  & \multicolumn{1}{|c|}{DAC} & $--$ & $--$ & $--$ & $--$ \\ \cline{2-6}
  & \multicolumn{1}{|c|}{LD} & $++$ & $++$ & $++$ & $++$ \\ \cline{2-6}
  & \multicolumn{1}{|c|}{MPC} & $--$ & $--$ & $--$ & $--$ \\ \cline{2-6}
  & \multicolumn{1}{|c|}{LCOM} & $--$ & $--$ & $--$ & $--$ \\ \cline{2-6}
  & \multicolumn{1}{|c|}{ILCOM} & $--$ & $--$ & $--$ & $--$ \\ \cline{2-6}
  & \multicolumn{1}{|c|}{TCC} & $++$ & $++$ & $++$ & $++$ \\ \hline
  \multirow{3}{*}{\rotatebox[origin=c]{90}{Design}} & \multicolumn{1}{|c|}{LOD} & $--$ & $--$ & - & $--$ \\ \cline{2-6}
   & \multicolumn{1}{|c|}{LEN} & $--$ & $--$ & $--$ & $--$ \\ \cline{2-6}
   & \multicolumn{1}{|c|}{CYC} & $--$ & $--$ & $--$ & $--$ \\ \cline{2-6}
\end{tabular}
\end{table}

Metric values can be extracted using the VizzMaintenance \cite{11} Eclipse plugin, which can extract values for the following class-level metrics: lines of code (LOC), number of attributes and methods (NAM), number of methods (NOM), weighted method count (WMC), response for class (RFC), depth of inheritance tree (DIT), number of children (NOC), coupling between objects (CBO), data abstraction coupling (DAC), locality of data (LD), message pass coupling (MPC), lack of cohesion in methods (LCOM) and its improved variant (ILCOM), tight class cohesion (TCC), lack of documentation (LOD), length of names (LEN) and number of classes in cycle (CYC). The aim is to provide a comprehensive view of object-oriented systems, capture their size, complexity, architecture and structure. Formal definitions for these metrics are available in the Compendium \cite{12} as well as detailed in existing research \cite{1,4}. 

The {ARiSA} model for maintainability is based on metric values aggregated according to the weights shown in Table \ref{tab:ARISAMetrics}. For each class, the percentage of extreme metric values is computed. A value is considered extreme when it is within the top or bottom 15\% of the metric's value range for all system classes. Once calculated for each metric, these are aggregated for each criteria in accordance with the weight and direction illustrated in Table \ref{tab:ARISAMetrics}. This results in a value between 0 (no extreme values) and 1 (all metric values are extremes) for each of the four criteria, which is then averaged to obtain the maintainability score of the class. For example, let us assume a class having a single extreme value, that of LOC. According to the information in Table \ref{tab:ARISAMetrics}, its Analyzability score is $\frac{2}{33}$. The numerator is given by the weight LOC has in Analyzability, represented by the two minus signs. The sum of the weights for the Analyzability criteria is 33. It follows then that Changeability is $\frac{2}{34}$, Stability is $\frac{1}{26}$ and Testability is $\frac{2}{33}$. The maintainability score of the class will be $\frac{\frac{2}{33}+\frac{2}{34}+\frac{1}{26}+\frac{2}{33}}{4}\approx0.0546$, or \textit{5.46\%}.

The ARiSA model was developed according to ISO 9126. While superseded by the newer ISO 25010, their differences regarding maintainability are small, so we believe models built on the older standard remain valid. When compared to the MI, the ARiSA model appears more comprehensive by taking into consideration object-oriented metrics that express cohesion, coupling and inheritance. In order to scale the model from class to system level, we take a similar approach to the MI. We calculate the maintainability score for each system class and calculate the geometric mean across system classes.

\section{\uppercase{Case Study}}
\label{sec:CaseStudy}
\subsection{Research Objective}
Our case study is organized according to the structure proposed by \cite{17}. The present section details case study design, while target applications and the procedure for data collection are discussed in Section \ref{sec:targetapp}. We report the results of our analysis in Section \ref{sec:analysis}, and address the threats to validity in Section \ref{sec:threats}, using the structured approach proposed by \cite{17}. 

The main objective of our work can be stated using the goal question metric approach \cite{18} as follows: \textit{"study the long-term evolution of maintainability in open-source software through the lens of metric-based software quality models"}. We distill the main objective into five research questions that will guide the design and analysis phases of the case study:

\bm{$RQ_{1}$}: \textit{Is maintainability correlated with software size?} The purpose of $RQ_{1}$ is twofold. First, we check that the proposed quality models do not actually measure software size. Second, we aim to study the influence of software size on reported maintainability. For instance, as LOC is one of the components of the MI, we expect it to be more readily influenced than the technical debt ratio. We also aim to address the na\"ive expectation that increased application size leads to lowered maintainability.

\bm{$RQ_{2}$}: \textit{How does the maintainability of open-source software change from one version to another?} We use the quality models detailed within the previous section, coupled with manual examination of the source code to identify both short and long-term trends during application development. This enables data triangulation as suggested by \cite{17}, lowering the risk of maintainability changes not being detected by the proposed models, and enables comparing results. We study the relation between software functionalities, software size and reported maintainability. We are also interested to identify and determine the impact that major changes to software, such as the addition of important features or refactoring, have on maintainability.

\bm{$RQ_{3}$}: \textit{Do differences exist between the maintainability scores reported for the same software version?} Existing research \cite{5} has shown this as an important possibility. Given that each quality model uses its own scale, we identify those versions where differences exist in the maintainability change between the last and current studied version.  

\bm{$RQ_{4}$}: \textit{What causes the differences observed within $RQ_{3}$?} We expect that answering $RQ_{4}$ will improve the understanding of the strengths and limitations the proposed models have when applied to complex software. In addition, we aim to study any versions where changes to maintainability were not discovered by any of the proposed quality models.

\bm{$RQ_{5}$}: \textit{Which of the proposed models is the most accurate at reflecting changes in software maintainability?} Answers to the previous research questions, especially $RQ_{4}$ will highlight the strengths and weaknesses of each proposed model. The purpose of $RQ_{5}$ is not to determine which model provides the best approximation of software maintainability, but to determine which model is most suitable at reflecting the changes that occur during the development of software. In addition, we aim to identify the proper context where each model provides valuable insight.

\subsection{Target Applications}
\label{sec:targetapp}
We defined a number of selection criteria for target applications. The goal was to ensure that our case study captures the development history for several applications, that results can be compared across the included systems and that results can be extrapolated to an important class of software systems. 

\begin{figure*}[!ht]
    \captionsetup[subfigure]{labelformat=empty,justification=centering}
    \centering
    \begin{subfigure}[b]{\textwidth}
        \includegraphics[width=\textwidth]{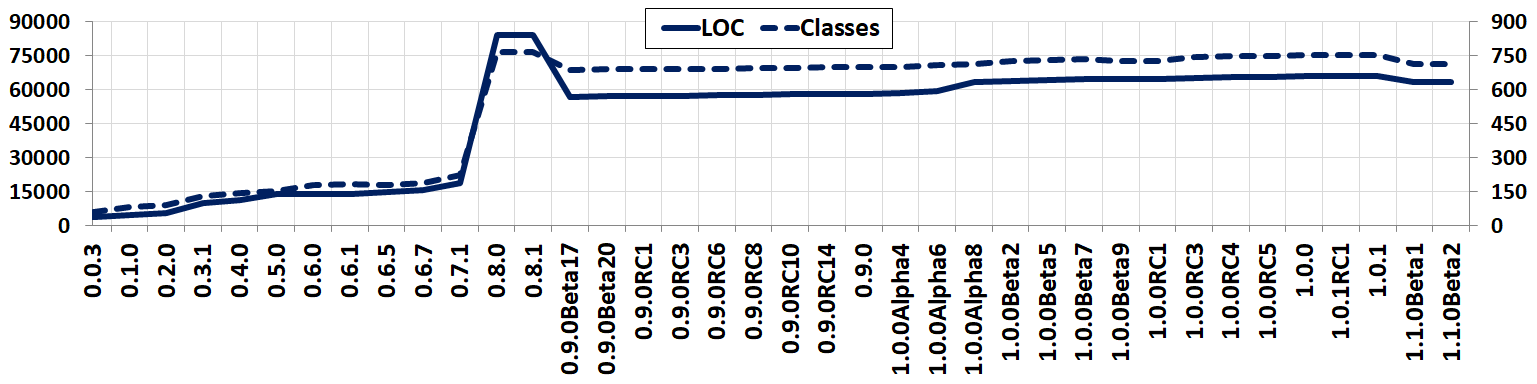}
    \end{subfigure}
    
    \begin{subfigure}[b]{\textwidth}
        \includegraphics[width=\textwidth]{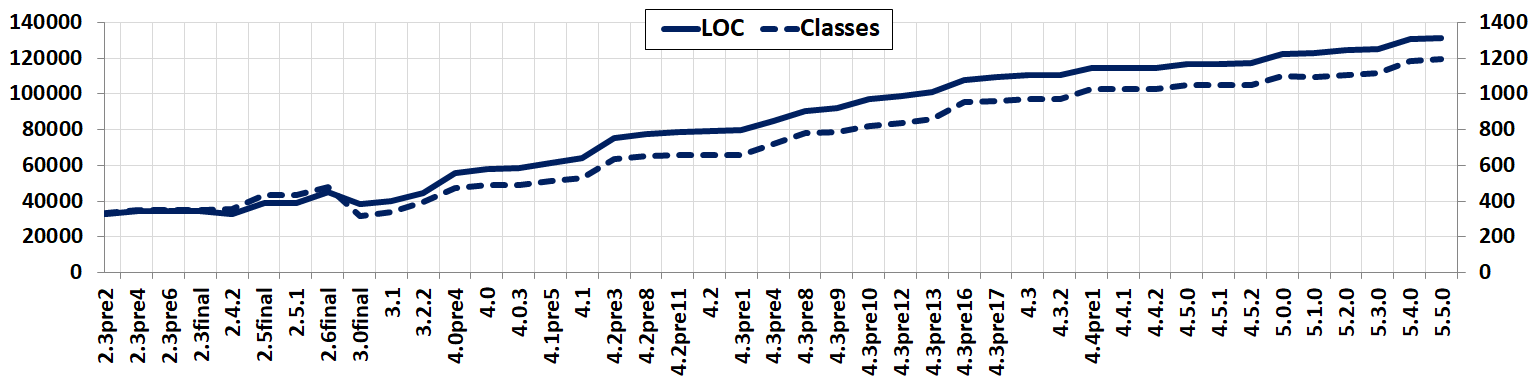}
    \end{subfigure}
    \par\medskip
    \begin{subfigure}[b]{\textwidth}
        \includegraphics[width=\textwidth]{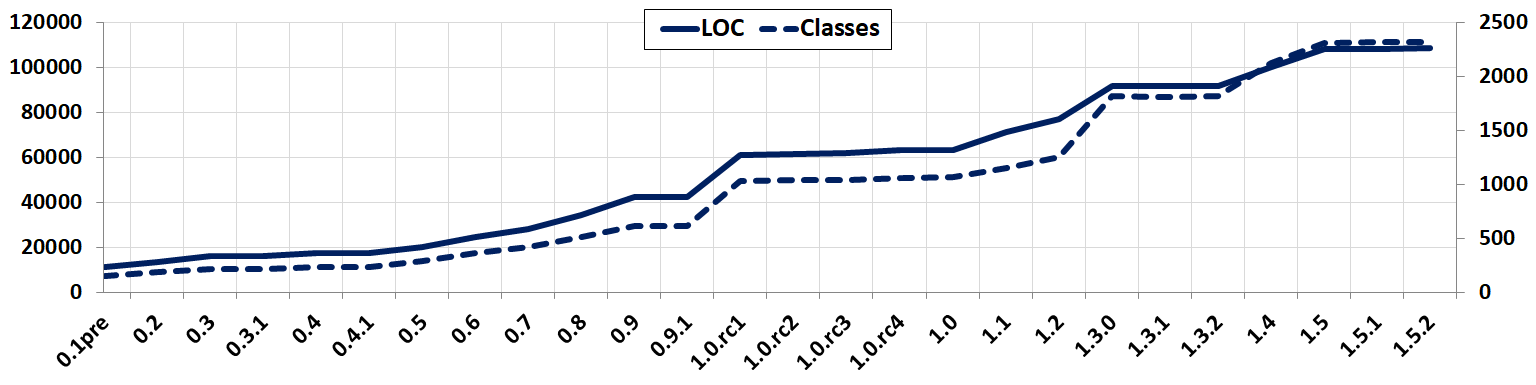}
    \end{subfigure}
    \caption{The size of FreeMind (top), jEdit (middle) and TuxGuitar (bottom) versions measured as lines of code (LOC) and number of classes.}
    \label{fig:ClassLOC}
\end{figure*}

First, we were interested in mature applications having an important user base, for which we had access to the entire development history. Second, we limited our selection to GUI-driven Java applications. This ensures the same code metrics and measurements can be used and compared across applications. Restricting the study to one popular application category allows producing a more representative picture. This is especially important when considering open-source software, as existing research \cite{1} shows that in many cases, development effort is not consistent and repositories include incomplete source code, that does not run or cannot be compiled. Finally, we steered towards applications having few dependencies to external libraries or other components, databases or the Internet. This facilitated setting up each application version and ensured that the source code for application functionality was self-contained.

We selected three open-source Java applications having long development histories, a well-established user base and a development repository that includes source code, release version history and associated change logs. The three applications are the FreeMind\footnote{http://freemind.sourceforge.net/wiki/index.php/Main\_Page} mind mapper, the jEdit\footnote{http://jedit.org} text editor and the TuxGuitar\footnote{http://www.tuxguitar.com.ar} tablature editor. 

\textbf{FreeMind} is a functionally-rich application for creating and editing mind maps that was previously used in empirical software engineering research \cite{2}. It has a consistent user base, with over 744k application downloads over the last 12 months, and over 24 million over its lifetime\footnote{All data points from https://sourceforge.net/, recorded December $12^{th}$, 2019}. Among the applications in our case study, FreeMind's development history can be traced closest to the project's start. The first released version was 0.0.3\footnote{We employ the version numbers provided by the developers}, and consisted of 3.700 lines of code (LOC) and 60 classes. Early versions do not have many functionalities, with frequent changes to source code structure and application architecture. Figure \ref{fig:ClassLOC} illustrates the evolution of LOC and class counts for each of the released versions. Early versions become more complex as more functionalities are integrated. An important jump is recorded in version 0.8.0, reflected both in the application's look and available functionalities as well as recorded metrics. Application code is refactored in version 0.9.0Beta17, resulting in reduced numbers for LOC and classes, without impacting application functionalities. Values for these metrics stabilize after this version, showing a slow increase until the most recent release. 

\textbf{jEdit} is a popular text-editor with plugin support and a large user base. The application recorded 99k downloads during the last year, and over 9 million over its development history. The first version was released in January 2000. Compared to other early application versions in our study, early versions of jEdit are the most stable and fully featured. The application was also used in empirical software research targeting the generation of GUI test cases \cite{2,3}, software metrics \cite{4} and maintainability \cite{5}. Figure \ref{fig:ClassLOC} shows a steady increase in the application code base throughout its development history. The single notable exception is version 3.0final, where refactoring caused a decrease in application size. Like FreeMind, plugin source code was not included in our evaluation. 

\textbf{TuxGuitar} is a multi-track guitar tablature editor. It employs an SWT-driven GUI across all recorded versions, while FreeMind and jEdit use AWT/Swing components. One of the application's key functionalities regards support for importing and exporting data in multiple formats. This is implemented in the form of plugins that ship with the default application installation and was included in our evaluation. 

Just like FreeMind and jEdit, TuxGuitar is also a popular application with an established user base, having over 218k downloads in the last year and 6.7 million during its lifetime. Figure \ref{fig:ClassLOC} illustrates the system's evolution in size, which we find to be similar to jEdit. The most significant difference is that we could not find instances of decreasing application complexity. However, it is possible that such changes were carried out between releases and compensated by new code.

Our study includes the 38 released versions of FreeMind, 43 in the case of jEdit and 26 TuxGuitar releases. As shows in previous research, we cannot make assumptions on source code in open-source repositories. In their study, \cite{1} found that one third of the applications considered in their case study required manual fixes before they could be started. As such, we imported each version into an IDE. The source code for each version was manually examined. For the older application versions, small changes had to be made to ensure they could be compiled and run under the Java 8 platform. An example of such changes regards variables named \textit{enum} in older source code. As Java 1.5 introduced the \textit{enum} keyword, these variables had to be renamed in order to ensure successful compilation. In several cases, open-source library code was found bundled together with application code. We separated this code into library files in the application classpath, to make sure it did not skew recorded metric values. We started each application version and checked every functionality to make sure the compiled source code was complete.

\subsection{Analysis}
\label{sec:analysis}
In this section we present our analysis of the collected data and provide answers to the research questions. To answer $RQ_{1}$, we consider the number of a system's classes as a suitable proxy for its size. While this does not account for differences in software architecture, concurrency or distributed software, we believe it to be suitable when limited to the selected category of applications. We calculate the Spearman rank correlation between the maintainability reported for each application version and its number of classes. Improved maintainability is represented by higher values of the MI, and lower values in the ARiSA model and technical debt ratio. Table \ref{tab:MaintainabilitySizeCorrelation} does not reveal strong correlation between any of the applications and quality models. 

FreeMind and jEdit results are similar. Both ARiSA and MI measurements are inversely-correlated with application size, but for different reasons. For the ARiSA model, classes added to later system versions serve to keep the mean value close to constant, even when application size increases. In the case of the MI, we observe that increased system size generally drives joint increases in Halstead volume, statement count and cyclomatic complexity. This illustrates the na\"ive expectation that lower maintainability is reported for larger applications. However, technical debt ratio remains independent of both application's sizes.

\begin{table}[t]
\caption{Spearman rank correlation between software size and maintainability measured using the ARiSA model, technical debt ratio (TDR) and the Maintainability Index (MI)}\label{tab:MaintainabilitySizeCorrelation} \centering
\begin{tabular}{|c|c|c|c|}
  \hline
  Application & ARiSA & TDR & MI \\ \hline
  FreeMind & -0.49 & 0.09 & -0.67 \\ \hline
  jEdit & -0.50 & 0.04 & -0.32 \\ \hline
  TuxGuitar & 0.67 & -0.71 & 0.64 \\ \hline
\end{tabular}
\end{table}

\begin{figure*}
  \includegraphics[width=\linewidth]{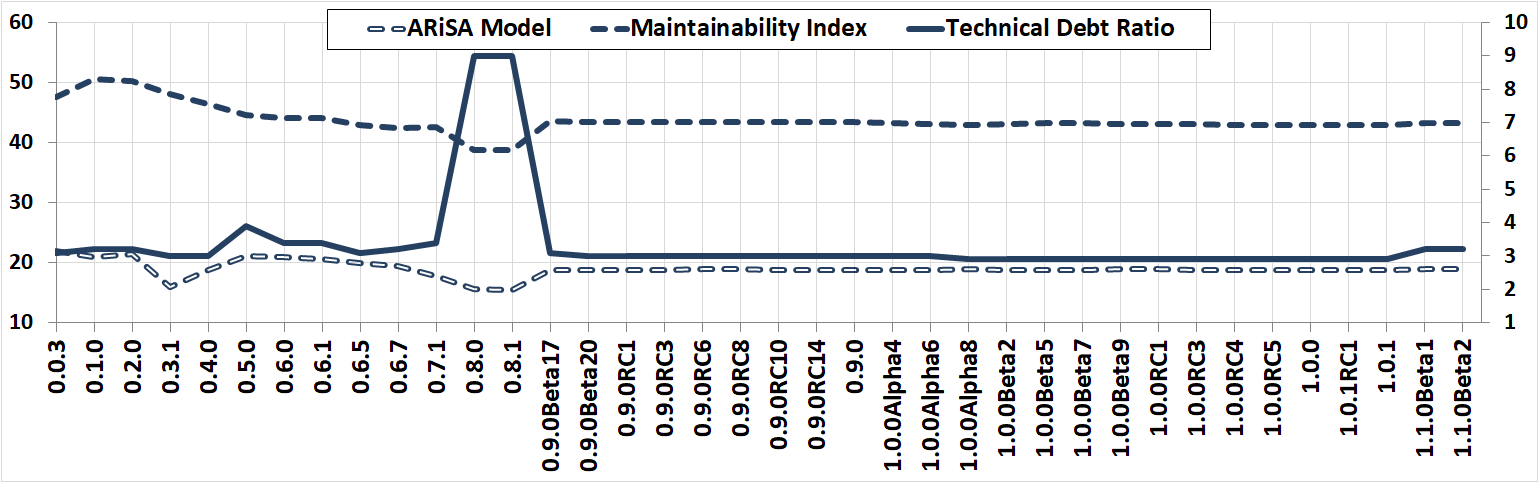}
  \caption{FreeMind maintainability according to the quality models. Technical debt ratio uses the scale on the right side.}
  \label{fig:FreeMindMaintainability}
\end{figure*}

\begin{figure*}
  \includegraphics[width=\linewidth]{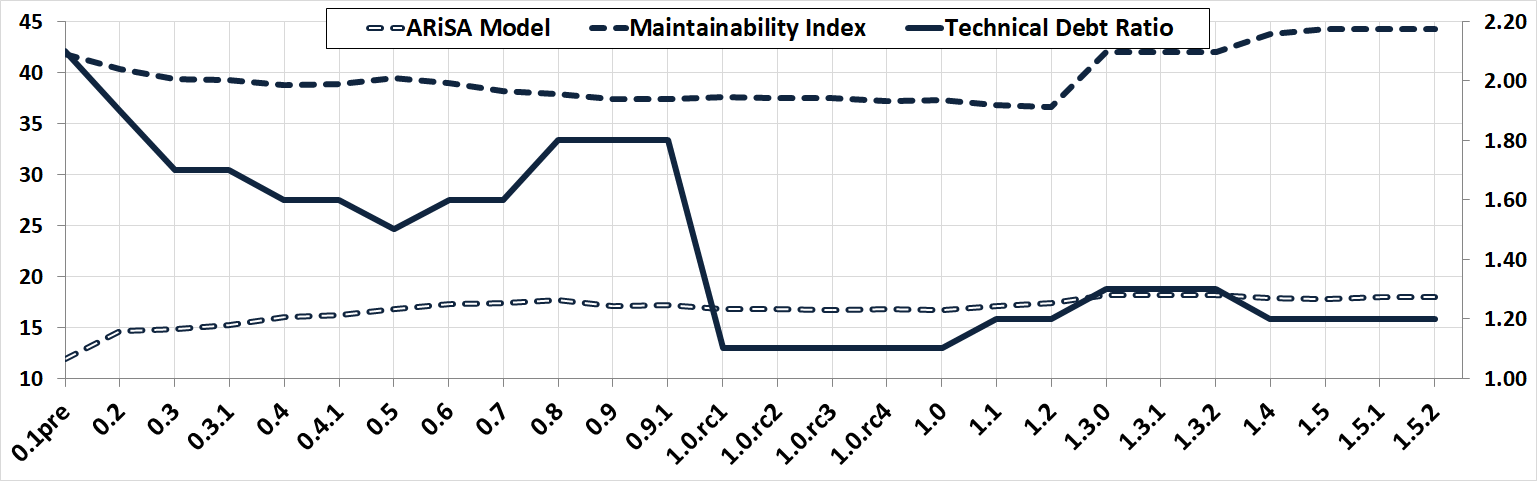}
  \caption{TuxGuitar maintainability according to the quality models. Technical debt ratio uses the scale on the right side.}
  \label{fig:TuxGuitarMaintainability}
\end{figure*}

TuxGuitar results are determined by a conscientious effort to keep technical debt in check within the project's latter versions. Figure \ref{fig:ClassLOC} shows important increases in system size for several versions, including 1.0rc1, 1.3 and 1.5. However, Figure \ref{fig:TuxGuitarMaintainability} shows very low levels of technical debt ratio. This is illustrated through the correlation values in Table \ref{tab:MaintainabilitySizeCorrelation}, as well as in Figure \ref{fig:TuxGuitarMaintainability}, which shows improved MI values after version 1.2 and a very low level of technical debt in all versions following 0.9.1.

\begin{figure*}
  \includegraphics[width=\linewidth]{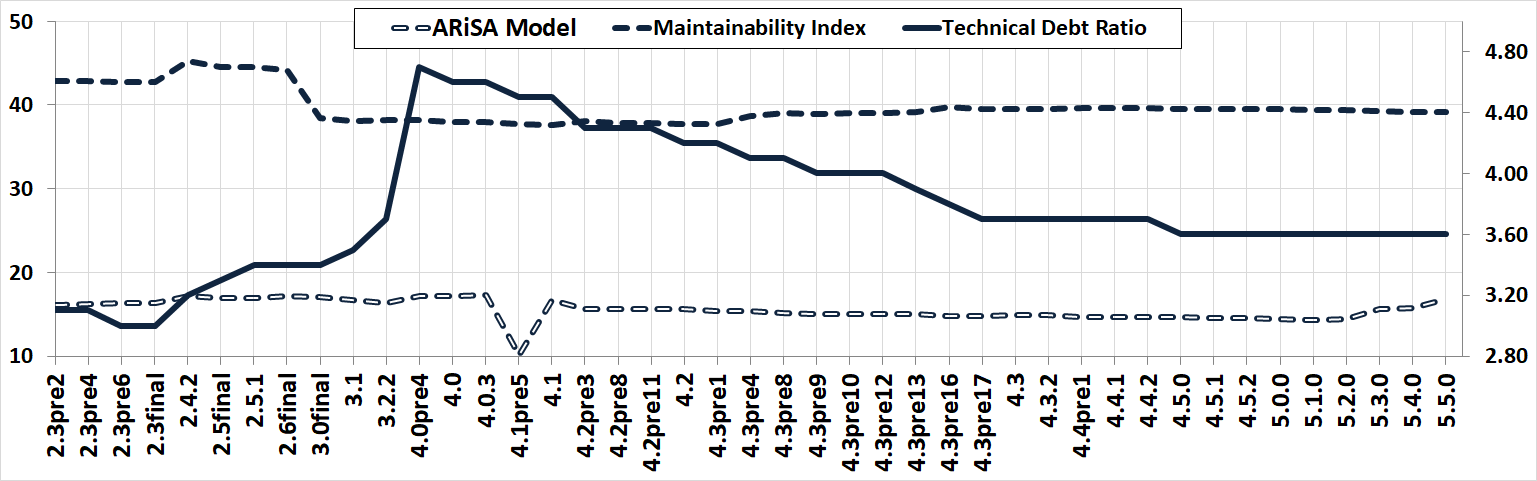}
  \caption{jEdit maintainability according to the quality models. Technical debt ratio uses the scale on the right side.}
  \label{fig:jEditMaintainability}
\end{figure*}

We answer $RQ_{2}$ using the maintainability data points in Figures \ref{fig:FreeMindMaintainability}, \ref{fig:jEditMaintainability} and \ref{fig:TuxGuitarMaintainability}. Data points were collected after normalizing values to the [0, 100] range\footnote{We used the $MI_{normalized}$ formula, while the numerical values from the ARiSA model and technical debt ratio represent percentages.}. Improved maintainability is illustrated using higher values of the MI, and lower values using the ARiSA model and technical debt ratio.

Our first observation continues the idea behind $RQ_{1}$. FreeMind versions newer than 0.9.0Beta17 have almost constant maintainability across all quality models, even with new features that increase class count and LOC by 20\%. For jEdit, the ARiSA-based model and MI see small changes between versions 4.2pre3 and 5.2.0, even with the increased application size shown in Figure \ref{fig:ClassLOC}. After version 4.0pre4, technical debt ratio remains on a downward slope, due to newly added code introducing few code smells and little technical debt. The same holds for TuxGuitar. Versions after 0.9.1 show improved maintainability, even if later application versions more than double LOC and class count, as shown in Figure \ref{fig:ClassLOC}.

The second observation regards early application versions. The smallest application versions in our case study are the early version of FreeMind, for which we record frequent changes to maintainability. The same can be observed for jEdit and TuxGuitar, and manual examination of the code shows that application structure remains fluid, with each release combining new features with refactoring. As expected, latter versions have a well defined architecture that does not change as much, with new features implemented over the existing system of plugins or event-handling systems.

In several cases, sequences with little variance in maintainability are bounded by versions where recorded values change abruptly. The most representative examples are FreeMind versions 0.8.0 and 0.9.0Beta17. All three models reflect the change in maintainability. Versions 0.8.* are the only ones to receive a \textit{B} maintainability rating according to their technical debt ratio, all the rest being \textit{A}'s. The root cause is a large push in development resulting in a sharp increase in LOC and class count. 0.8.0 is the first version to use external libraries for XML processing and input forms. During use, it is clear that version 0.8.0 is more complex and fully-featured, with many visible changes at the user interface level, including more complex application preferences and features for mind map and node management. The scope of changes remains apparent at source file level, with only 21 out of the 92 source files remaining unchanged from version 0.7.1. The number of source files also increases greatly, from 92 to 469. Much of the observed discrepancy between numbers of source files, classes and LOC between the versions can be explained by the newer application including 272 classes that were generated by the JAXB libraries encoding most of the actions that can be performed within the application. These classes contributed with 49.434 lines to the witnessed inflation of LOC. These classes are removed in version 0.9.0Beta17, which consists of only 127 source files. However, much of the maintainability variance is not a result of changes to system size. Version 0.9.0Beta17 has similar maintainability to 0.7.1, but tripled the class count and LOC. 

In the case of jEdit 4.0pre4, our examination reveals a disproportionate increase of the technical debt when compared to LOC. This is caused by adding high cyclomatic complexity code for text area management, text buffer events and the application document model. Across most jEdit versions, these functionalities include the application's most complex code.

TuxGuitar is characterized by a very good ratio of technical debt. Version 1.0rc1 sees technical debt ratio drop further, and keep the best values recorded in our case study. This is due to extensive refactoring that eliminated many code smells and lowered total technical debt from 326 hours to 290 hours. This, coupled with increased size caused by adding support for the song collection browser and new plugins leads to the observed result. The MI also records an improvement in version 1.3.0. Manual examination reveals it to be only apparent, as it is caused by the addition of over 930 classes describing custom application actions. They have few statements and low cyclomatic complexity which skews the MI by decreasing mean values.

In order to address $RQ_{3}$ we identify software versions where the proposed quality models are in disagreement. We take into consideration that each quality model uses its own scale, which does not allow us a direct comparison of the results. We feel that a good example are FreeMind versions 0.8.*, where a sharp increase in technical debt ratio appears together with a decrease of the MI. However, even these version are rated \textit{B} on SonarQube's maintainability scale, which ranges from \textit{A} (best) to \textit{E} (worst). As such, these versions do not fall under the purview of $RQ_{3}$. Analyzing the maintainability data reveals quality models are in disagreement for jEdit version 4.0pre4, 4.1pre5 and TuxGuitar versions 1.0rc1 and 1.3.0. 

We examine these versions in detail in order to provide an answer for $RQ_{4}$. Version 4.0pre4 is an important update, where all source files are modified, 80 new classes are added and LOC increases by 25\%. However, the complexity of added code does not differ from that of the previous version's code base, it is averaged out from the MI and the ARiSA model. However, SonarQube metrics report over 2700 new code smells, which raise technical debt by 50\%, leading to decreased maintainability. jEdit version 4.1pre5 highlights one limitation of the ARiSA model, which examines class-level maintainability in the context of the entire system. In this case, the observed improvement is the result of the way maintainability is calculated. For instance, the ARiSA model reports different maintainability levels for some of the most complex system classes, even when their code was not changed.

TuxGuitar 1.0rc1 is an important update, during which all but 3 source files were modified. Both LOC and class count increased by 50\%. New functionalities, such as the song collection browser and data management plugins were added. Like in the previous case, newly introduced code shows similar metrics to the existing code base, so changes do not influence the MI or ARiSA model. However, as detailed in the $RQ_{2}$ paragraph, technical debt is decreased, both in absolute (fewer code smells and less technical debt) and relative size.

In the case of version 1.3.0, we again find improved MI to be the result of adding a large number of small, low-complexity classes that skew mean values, and not the result of a real improvement to software quality. Added code also causes a small increase in technical debt ratio.

In order to provide an answer for $RQ_{5}$, we examined all consecutive application version pairs manually. We started by examining each version's package structure, followed by a perusal of the source code in new packages. For packages found in several versions, we examined the code changes at class-level in order to localise architecture details and application features. Our analysis confirmed the findings reported for $RQ_{3}$ and $RQ_{4}$. As such, it is our belief that from a software development point of view, the technical debt ratio is the most accurate of the examined maintainability models. We also found it to be suitable for discovering maintainability changes across versions, as evidenced when examining jEdit 4.0pre4 and TuxGuitar 1.3.0. 

In contrast, we believe the MI is useful when deployed for smaller code units. At method or class-level, the skewing effect of data averaging is not present, so the MI provides useful indication regarding code complexity and readability. The scale proposed in \cite{32} can be used during development and code review to prevent issues from accumulating. We found this to be important especially during an application's early versions. As these versions already contain some of the most complex code, preventing them from being part of the technical debt package can be a worthwhile goal in the long term.

The ARiSA model is best used to discover maintainability hot spots within the code base. Unlike the MI, it employs a wide range of metrics, making it sensitive to object-oriented code. However, using lower-level code metrics makes it less amenable to providing a general assessment of an application. Finally, we also observe that both MI and the ARiSA models can be readily adapted or implemented for new languages or platforms, while doing so for the technical debt ratio requires significant more work.

\subsection{Threats to Validity}
\label{sec:threats}
The case study was guided by \cite{17}, and consisted of the following steps: identifying and preparing target applications, manual source code examination, data collection and analysis. We documented all steps to ensure the study can be replicated.

Target application selection targeted complex, popular applications for which we could analyze both early as well as mature versions. We restricted candidate applications to a narrow, but nevertheless popular type to allow data triangulation and comparison. To the best of our knowledge, target applications were developed by independent teams. The study's authors had no involvement with the development of the target applications, or of the metric extraction tools. Given that each extraction tool uses its own metric definitions, this can impact obtained results \cite{33,34}. We employed several tools to extract metrics values\footnote{MetricsReloaded IntelliJ plugin and the VizzMaintainance Eclipse plugin} and cross-checked the accuracy of collected data, which is available on our website\footnote{https://bitbucket.org/guiresearch/tools}.

The quality models employed were selected according to their presence in existing literature, varying level of implementation complexity and the possibility of obtaining a numerical value to represent maintainability for each software version. 

We calculated both line and statement count versions of the Maintainability Index. As we found they correlated almost perfectly (coefficient over 0.99), we used the statement-based version. For the ARiSA model, we aggregated values using the geometric mean. While a different aggregation strategy might improve the accuracy of the model, researching this remained beyond the scope of the present paper. Before the research questions were addressed, we calculated the correlation between the reported values, and made sure there was no risk of collinearity. 

Threats to internal validity were addressed by carrying out the manual source code examination before data analysis. This was done to prevent biases that could be introduced by the availability of maintainability data. Code examination was assisted by scripts that allowed us to easily compare the source code across different versions and summarize the size and scope of changes. 

External validity was addressed through the target application selection process. Our analysis facilitated finding trends that were common across the applications. Added value for cementing these results is possible through a replication study with a larger scope. However, the requirements for testing that each version is complete and sanitizing it, as described in Section \ref{sec:targetapp} make this a laborious process.

\section{\uppercase{Related Work}}
\label{sec:RelatedWork}
Existing studies regarding maintainability can be classified in two categories. The first is where maintainability is computed using software metrics and the second is where maintainability is primarily considered a software quality factor.

The relation between maintainability and software metrics has been thoroughly investigated \cite{8,22,21,5}. This is especially true after the introduction of the Chidamber \& Kemerer metrics suite \cite{7}, as research highlighted that predicting software maintenance effort was possible \cite{22}. Research also showed that evaluating maintainability should take cohesion and coupling, as viewed within the lens of object-oriented systems into account \cite{25}. During our literature review we identified two important drawbacks of existing approaches. First, most of the studies prove the influence of several factors over maintainability, but lack a computational or methodological approach that can be used in practice. Second, where they exist, numerical threshold values for maintainability, such as presented in Section \ref{sec:MI} remain debatable.

The most comprehensive survey of the relation between metrics and software quality attributes, including maintainability, can be found in \cite{12}. The survey is built on the ISO 9126 quality model. The approach is based on decomposing quality characteristics in sub-characteristics and computing the influence of several software metrics on them, proposing a descriptive method. This represents the structured foundation for the ARiSA model detailed in Section \ref{fig:ARISAModel}.

An approach to evaluate maintainability based on the most recent ISO 25010 standard is proposed in \cite{26}. The study offers an interesting perspective, but it is still in early stages, and other direct or indirect measures should be included. The SIG Maintainability model offers a similar proposal \cite{27,8}, by associating source code properties to maintainability sub-factors, and validating this hypothesis on a large number of industrial applications. The result is represented using a scale of five levels of influence between sub-factors and code properties. The most important drawback of studying maintainability as a software quality factor is the fact that it is descriptive, establishing relations or influences, but lacks a computational perspective. This precludes such models from being included as automated steps of software development pipelines, and represents an important impediment to their adoption by practitioners.

\section{\uppercase{Conclusions}}
\label{sec:Conclusion}
We carried out a longitudinal exploratory case study where three software quality models and manual source code examination were used to study software maintainability over the long-term. We report the results of our analysis that covered over 18 years of development history for three complex, popular applications.

Our first conclusion, that we observed for all three target applications is that maintainability can remain independent of application size. While class count and LOC generally increased throughout application development, we noticed consistent refactoring efforts that improved reported maintainability. Furthermore, we observed source code that was newly added to latter versions to have very good maintainability, helping to keep overall software quality high. Determining the reason for this is beyond the scope of our research. However, we believe a well established application architecture, or having experienced developers contribute to complex application versions might contribute to this effect.

We also highlight the common development trend of these open-source applications. Early versions are susceptible to fluctuations in source code quality, as well as metric values \cite{4}. Once application architecture, in the form of package structure, used libraries and event handling mechanisms are established, we find smaller fluctuations in subsequent versions. An important exception to this is represented by the existence of milestone versions, which have a large variance in quality. These are characterized by the addition of many new features, which is also apparent while using the application. In the case of jEdit 4.0pre4 and TuxGuitar 1.0rc1, these are also evident from version numbers. However, the opposite is not true.

Regarding the quality models employed in our case study, we believe each of the methods can be useful in discovering maintainability changes. While beyond the scope of our present paper, all three models can be used with finer granularity, at package and class levels. Each model contributed insight regarding the quality of the target applications, and we believe that comprehensive analysis using several models is useful in order to improve the characterization of software quality.

We intend to expand our study to cover additional system types prevalent today, such as frameworks, distributed applications or those targeting mobile devices. We aim to further study system size and complexity beyond line of code and class counts in order to improve our understanding of its relation with maintainability. In addition, we believe a cross-sectional approach is also valuable, as it can improve our baseline by facilitating the study of a larger number of target applications.

\bibliographystyle{apalike}
{\small
\bibliography{references}}
\end{document}